\documentstyle[twocolumn]{mn}
\newcommand{\mincir}{\raise
-2.truept\hbox{\rlap{\hbox{$\sim$}}\raise5.truept
\hbox{$<$}\ }}
\newcommand{\magcir}{\raise
-2.truept\hbox{\rlap{\hbox{$\sim$}}\raise5.truept
\hbox{$>$}\ }}
\newcommand{\minmag}{\raise-2.truept\hbox{\rlap{\hbox{$<$}}\raise
6.truept\hbox
{$>$}\ }}
\newcommand{\be}{\begin{equation}}
\newcommand{\ee}{\end{equation}}
\newcommand{\ba}{\begin{eqnarray}}
\newcommand{\ea}{\end{eqnarray}}
\newcommand{\brr}{\begin{array}}
\newcommand{\nn}{\nonumber \\} 
\newcommand{\err}{\end{array}}
\newcommand{\bc}{\begin{center}}
\newcommand{\ec}{\end{center}}

\input{psfig.sty} 
\title{The Clustering of Lyman-break Galaxies}
\author[Coles et al.]
{Peter Coles$^{1}$, 
Francesco Lucchin$^{2}$, Sabino Matarrese$^{3}$ and
Lauro Moscardini$^{2}$ \\
$^1$Astronomy Unit, School of Mathematical Sciences,  
Queen Mary \& Westfield College, Mile End Road,  
London E1 4NS\\ 
$^2$Dipartimento di Astronomia, Universit\`a di Padova,
vicolo dell'Osservatorio 5, I--35122 Padova, Italy\\
$^3$Dipartimento di Fisica G. Galilei, 
Universit\`{a} di Padova, via Marzolo 8, I--35131 Padova, Italy}

\begin{document}

\maketitle

\begin{abstract}
We calculate the statistical clustering of Lyman-break galaxies predicted in a
selection of currently-fashionable structure formation scenarios. These models
are all based on the cold dark matter model, but vary in the amount of dark
matter, the initial perturbation spectrum, the background cosmology and in the
presence or absence of a cosmological constant term. If Lyman-break galaxies
form as a result of hierarchical merging, the amplitude of clustering depends
quite sensitively on the minimum halo mass that can host such a galaxy.
Interpretation of the recent observations by Giavalisco et al. (1998) would
therefore be considerably clarified by a direct determination of the relevant
halo properties. For a typical halo mass around $10^{11} h^{-1} M_\odot$ the
observations do not discriminate strongly between cosmological models, but if
the appropriate mass is larger, say $10^{12} ~h^{-1} M_\odot$ (which seems
likely on theoretical grounds), then the data strongly favour models with a low
matter-density. 
\end{abstract}


\section{Introduction}

Developments in observational techniques have recently led to an explosion of
interest in the properties of cosmological objects at such high redshifts that
the lookback time at which they are seen is a considerable fraction of the age
of the Universe. Such objects therefore furnish the opportunity to probe
directly the evolution of galaxy clustering, bridging the gap between local
observations of large-scale structure (i.e. observations with lookback times
that are small compared to the age of the Universe) and observations of the
cosmic microwave background radiation (where the lookback time is virtually
equal to the age of the Universe). 

In Matarrese et al. (1997; hereafter Paper I) and Moscardini et al. (1998;
hereafter Paper II), we discussed  high-redshift clustering phenomena from a
theoretical perspective in order to provide a general framework within which
these high-redshift phenomena can be interpreted. The formalism we derived can
be used to make detailed predictions of statistical measures of clustering in
specific cosmological scenarios and also  makes explicit the main sources of
theoretical  uncertainty in these predictions. This allows one to make a
realistic assessment of how models of structure formation fare in the face of
results from particular observational programmes. In Paper II we applied this
approach to a confrontation of different cosmological models with various
galaxy clustering data, including a brief discussion of the Lyman-break
galaxies (LBGs) presented in Steidel et al. (1998). 

The properties of LBGs are presently undergoing a great deal of scrutiny
because of the opportunities they present to study systematically the spatial
clustering of galaxies at extremely high redshifts. It is the purpose of this
paper to deploy the techniques of Papers I \& II in a systematic comparison of
a selection of  currently popular structure formation models with recent
results on the angular correlations of LBGs presented by Giavalisco et al.
(1998). 

LBGs are identified by using an efficient photometric technique that allows the
identification of candidate high-redshift objects through the shifting of their
Lyman limit cutoff into a particular colour filter (Steidel \& Hamilton 1993).
Recently Steidel et al. (1996, 1998) began a survey for $z\sim 3$ galaxies
using this Lyman-break technique by observing five different fields altogether
covering approximately 700 arcmin$^2$. They found 871 candidates with magnitude
${\cal R} \le 25.5$. The subsequent spectroscopic identification for 376
galaxies showed that approximately 90 per cent of the objects have $2.6\le z
\le 3.4$, with a median redshift $z=3.04$ and a small r.m.s. ($\sigma_z=0.27$).

The interpretation of LBGs is not entirely straightforward. They are certainly
small, but are highly luminous objects. Moreover, interpretating their strong
interstellar absorption lines as essentially due to gravitational motions,
their masses seem to be similar to those of local bright galaxies (Giavalisco,
Steidel \& Macchetto 1996). This favours the interpretation that they comprise
a massive halo within which the formation of a galaxy is in progress (see e.g.
Baugh et al. 1998). The alternative possibility that LBGs are relatively
low-mass ($\sim 10^{10} M_\odot$) objects with an intense starburst activity
has been discussed by Somerville (1997). 

The problem of the LBGs, and in particular the probability that a strong
concentration at $z\sim 3$, similar to that reported by Steidel et al. (1998),
arises in particular cosmological scenarios, have been discussed by various
authors by using both analytical and N-body techniques (Mo \& Fukugita 1996;
Jing \& Suto 1998; Governato et al. 1998; Bagla 1997a; Wechsler et al. 1997;
Peacock et al. 1998; Paper II). The resulting picture seems to indicate that at
$z\sim 3$ LBGs are strongly biased because they lie at the highest peaks of the
density field (Kaiser 1984). They are therefore expected to be strongly
clustered in all currently-fashionable cosmological models. If this
interpretation is correct, LBGs would be the progenitors of the population of
massive ellipticals (Steidel et al. 1998) or cluster galaxies we see at $z=0$
(Governato et al. 1998). 

The plan of this {\em Letter} is as follows. In Section 2 we describe our
formalism to study the clustering at high redshift. In Section 3 we present the
cosmological models used in the following analysis. The predictions of the
correlation functions of the LBGs are shown in Section 4. The final discussion
and conclusions are presented in Section 5. 

\section{Clustering at High Redshift}

In Papers I \& II we  showed that the observed spatial correlation function
$\xi_{\rm obs}$ in a given redshift interval ${\cal Z}$ is an appropriate
weighted average of the mass autocorrelation function $\xi$ with the mean
number of objects ${\cal N}$ and effective bias factor $b_{\rm eff}$, defined
below in equation (\ref{eq:b_eff}), in that range: 
\ba
 \xi_{\rm obs}(r) & = & N^{-2} \int_{\cal Z} d z_1 dz_2 \nn
 & \times &  
{\cal N}(z_1) ~{\cal N}(z_2) ~b_{\rm eff}(z_1) ~b_{\rm eff}(z_2)
~\xi(r,\bar z) \: , 
\label{eq:xifund}
\ea
where $N \equiv \int_{\cal Z} d z' {\cal N}(z')$ and $\bar z$ is an
intermediate redshift between $z_1$ and $z_2$; see Papers I \& II for details. 

The factor of $b_{\rm eff}$ which appears in equation (\ref{eq:xifund}) is a
consequence of our lack of understanding of the details of the galaxy formation
process and the consequently uncertain relationship between fluctuations in
matter density $\delta_{\rm m}$ and galaxy number-density $\delta_{\rm n}$. We
assume that objects with given intrinsic properties (such as mass $M$) and at
different redshifts $z$ can have different bias parameters, which we call
$b(M,z)$. For each set of objects, however, the bias is assumed still to be
linear; it is also {\em local} (e.g. Coles 1993), in the sense that the
propensity of galaxies to form at a given spatial location ${\bf x}$ depends
only on the matter density at that point: 
\be
\delta_{\rm n}({\bf x}; M,z) \simeq b(M,z) \delta_{\rm m}({\bf x},z);
\label{eq:bialoc}
\ee
so that no environmental or cooperative effects in galaxy formation (e.g. Babul
\& White 1991; Bower et al. 1993) are permitted. If we  assume such a bias
between the galaxy and mass fluctuations, the {\em effective} bias factor
$b_{\rm eff}(z)$ which appears in equation (\ref{eq:xifund}) can be expressed
as a suitable average of the ``monochromatic'' bias $b(M,z)$ (i.e. the bias
factor of each single object): 
\be 
b_{\rm eff}(z) \equiv {\cal N}(z)^{-1} \int_{\cal M} d\ln M' ~b(M',z) 
~{\cal N}(z,M')\; .
\label{eq:b_eff}
\ee
In principle the variable $M$ (and its range ${\cal M}$) stands for any
intrinsic properties of the object in question (e.g. mass, luminosity, etc.) on
which the selection of the object into an observational sample might depend.
From here on, however, we shall assume that all such properties can be reduced
to a dependence on the mass of the halo within which the object (galaxy) forms.
This assumption is to some extent debatable (see, e.g., Kauffmann, Nusser \&
Steinmetz 1997; Roukema et al. 1997) but seems a reasonable starting point in
the limited range of redshifts relevant to the LBG population. Moreover,
Haehnelt, Natarajan \& Rees (1997) have recently shown that a good fit to the
LBG luminosity function can be obtained by assuming a linear relation between
star formation rate and halo mass, which implies a constant ratio of
mass-to-UV-light. Henceforth in this study, therefore, $M$ can be taken to
stand for the mass of the parent halo of the LBG. 

In order to predict the clustering properties of objects as a function of $z$
we need to understand how the relationship between these objects and the
underlying mass distribution evolves. In most fashionable models of structure
formation the growth of structures on a given mass scale is driven by the
hierarchical merging of sub-units. One begins by calculating the bias parameter
$b(M,z)$ for haloes of mass $M$ and `formation redshift' $z_f$ at redshift
$z\leq z_f$ in a given cosmological model. The result is 
\be
b(M,z\vert z_f) = 1 + {D_+(z_f) \over \delta_c D_+(z)}
\biggl( {\delta_c^2 \over \sigma_M^2 D_+(z_f)^2 } - 1\biggr) \;, 
\label{eq:b_mono}
\ee
where $\sigma^2_M$ is the linear mass-variance averaged over the scale $M$
extrapolated to the present time ($z=0$) and $\delta_c$ is the critical linear
overdensity for spherical collapse [$\delta_c={\rm const}=1.686$ in the
Einstein-de Sitter case, while it depends slightly on $z$ for more general
cosmologies (Lilje 1992)]. The above expression originally appeared (in a 
slightly different form) as equation (6) of Cole \& Kaiser (1989). It was then
later discussed by Mo \& White (1996) who also compared it with the results of
numerical experiments which showed good agreement with the simple theoretical
form. The general non-linear relation between the halo and the mass density
contrast has been recently obtained by Catelan et al. (1998), by solving the
continuity equation for dark matter haloes. A complementary study by Bagla
(1997b) has further explored the clustering of haloes using numerical
experiments; see also Ogawa, Roukema \& Yamashita (1997). 

As in Paper I, we can estimate the effective bias by assuming that the objects
observed in a given survey represent all haloes exceeding a certain cutoff mass
$M_{\rm min}$ at any particular redshift. In other words, we assume that there
is a selection function $\phi(z,M)=\Theta(M-M_{\rm min})$ at any $z$, where
$\Theta(\cdot)$ is the Heaviside step function. In this way, by modelling the
linear bias at redshift $z$ for haloes of mass $M$ as in equation
(\ref{eq:b_mono}) and by weighting it with the theoretical mass--function $\bar
n(z, M)$ which we can self--consistently calculate using the Press-Schechter
(1974) theory, we can obtain the behaviour of $b_{\rm eff}(z)$ directly. If we
were to assume that rapid merging continued up to the present then this model
(that in Papers I \& II we called the {\em merging model}) is completely
defined by the initial amplitude of primordial density fluctuations, because
that determines the scale of non-linearity at each epoch $z$. This would mean
that the parameter $M_{\rm min}$ is fixed by requiring the present population
of galaxies to have been entirely produced by a merger-driven hierarchy.
Accordingly the present-day value of $b_{\rm eff}(z=0)$, which can be extracted
by comparing local measurements with the mass fluctuations predicted in a given
model, determines $M_{\rm min}$. 

We feel, however, that the assumption that one can compare the clustering of
present-day galaxies directly with that of LBGs (which are selected in an
entirely different way) is rather unsafe. Moreover, it may well be the case
that instantaneous merging is not a good approximation for the later stages of
clustering evolution. It is more reasonable therefore to regard $M_{\rm min}$
as a free parameter and not attempt to relate the properties of LBGs to local
galaxies. In Paper I \& II we called the model obtained by letting $M_{\rm
min}$ be a free parameter the  {\em transient model}, because the original
theoretical motivation for it was the case of high-$z$ QSOs which have no
obvious counterpart among the local galaxy population. It should, however, also
be a good model for LBGs. The choice of minimum halo mass for the LBG case is
not obvious, so in the following we give results for two representative cases
($10^{11}$ and $10^{12} h^{-1} {\rm M}_{\odot}$). The higher of these values is
favoured by the properties of absorption lines of these objects (Steidel et al.
1998; see, however, Somerville 1997). 

The computation of clustering properties using equation (1) is completed by the
specification of the matter covariance function and its evolution with $z$,
i.e. $\xi(r, z)$. As in Papers I \& II we use a method based on the original
suggestion by Hamilton et al. (1991) and developed by Peacock \& Dodds (1994),
Jain, Mo \& White (1995) and Peacock \& Dodds (1996) to calculate the evolution
of perturbations into the non-linear regime. This technique also takes account
of different background cosmologies, possible contributions from a cosmological
constant and can be applied to a variety of initial perturbation spectra. 
 
\section{A Suite of Trial Cosmologies} 

In this paper we consider a set of cosmological models which can all be
regarded as variations on the basic cold dark matter (CDM) scenario. Although
the Standard CDM model (SCDM) is no longer regarded as a good fit to
observations of galaxy clustering and the microwave background, there are
several alternatives with many of the same basic features but with differences
in detail. In a general way, the initial (linear regime) power spectrum for all
these models, which provides the initial conditions for the clustering
evolution calculations discussed above, can be represented by $P_{\rm
lin}(k,0)=P_0 k^n T^2(k)$, where we use the fitting formula of the CDM transfer
function $T(k)$ as given by Bardeen et al. (1986). To fix the amplitude of the
power spectrum (generally parametrised in terms of $\sigma_8$, the r.m.s.
fluctuation amplitude inside a sphere of $8 h^{-1}$ Mpc) we either attempt to
fit the present-day cluster abundance or the level of fluctuations observed by
COBE (Bunn \& White 1997). 

We will consider the following specific models: the {\bf SCDM} model, as
reference model, with $n=1$ and a normalization consistent with the COBE data
($\sigma_8=1.22$); a different version of the SCDM model (hereafter called {\bf
SCDM$_{CL}$}) with a reduced normalization ($\sigma_8=0.52$) producing a
cluster abundance in agreement with the observational data (Eke, Cole \& Frenk
1996; see also Viana \& Liddle 1996); a COBE-normalized tilted model (hereafter
{\bf TCDM}; see e.g. Lucchin \& Matarrese 1985) with $n=0.8$, $\sigma_8=0.72$
and high baryonic content (10 per cent; see White et al. 1996; Gheller, Pantano
\& Moscardini 1998); a different version of the previous model, hereafter {\bf
TCDM$_{GW}$}, with a reduced normalization of the scalar perturbations
($\sigma_8=0.51$) taking into account the possible production of gravitational
waves, as predicted by some inflationary theories (e.g. Lucchin, Matarrese \&
Mollerach 1992; Lidsey \& Coles 1992); an open CDM model (hereafter {\bf
OCDM}), with a matter density parameter $\Omega_{0m}=0.4$, a Hubble parameter
$h=0.65$ and COBE--normalized ($\sigma_8=0.64$); a low--density CDM model
(hereafter {\bf $\Lambda$CDM}) always with $\Omega_{0m}=0.4$ but with a flat
geometry provided by the cosmological constant, with $h=0.65$ and
COBE--normalized ($\sigma_8=1.07$). 

\section{Predictions and Observations} 

We presented calculations of the spatial correlations of LBGs in Paper II.
Although the recent paper by Giavalisco et al. (1998) included an updated form
of the redshift distribution, we have verified that this does not alter the
predictions significantly. Observational estimates of the spatial two-point
function are not yet available but Giavalisco et al. (1998) have presented
results for the angular correlations of LBGs. In our previous papers we showed
that, in the {\em small--angle} approximation, the observed two-point angular
correlation function $\omega_{\rm obs}$ can be expressed as: 
\ba
 \omega_{\rm obs}(\vartheta)  & = & N^{-2} 
\int_{\cal Z} d z ~G(z) ~{\cal N}^2(z) ~b^2_{\rm eff}(z)\nn
 & & \times \int_{-\infty}^\infty
d u ~\xi[r(u,\vartheta,z) , z] \;, 
\ea
where $r(u,\vartheta,z) \equiv a_0 \sqrt{u^2 + x^2(z) \vartheta^2}$ and $x(z)$
depends on the background cosmology; see Paper II for details. Note that the
redshift distribution $N(z)$ is an observed quantity for a given survey so we
do not need to specify this function theoretically. In the following analysis
we adopt the same $N(z)$ presented in Giavalisco et al. (1998). 

In Figure 1 we compare the predictions of $\omega_{\rm obs}(\vartheta)$ for
each of the models presented in the previous section and in each case for two
choices of the minimum halo mass with the observational results obtained by
Giavalisco et al. (1998) using the 871 LBG candidates. It can be seen that most
models are consistent with the observations if the minimum halo mass is
relatively small ($10^{11} h^{-1} M_\odot$), whereas the larger choice is much
more strongly constrained. Indeed in the latter case only the two low-density
models $\Lambda$CDM and OCDM are consistent with observations. The SCDM model
does fit the results at larger angular scales, but is discrepant on small
scales. 

\begin{figure}                  
\centerline{\psfig{file=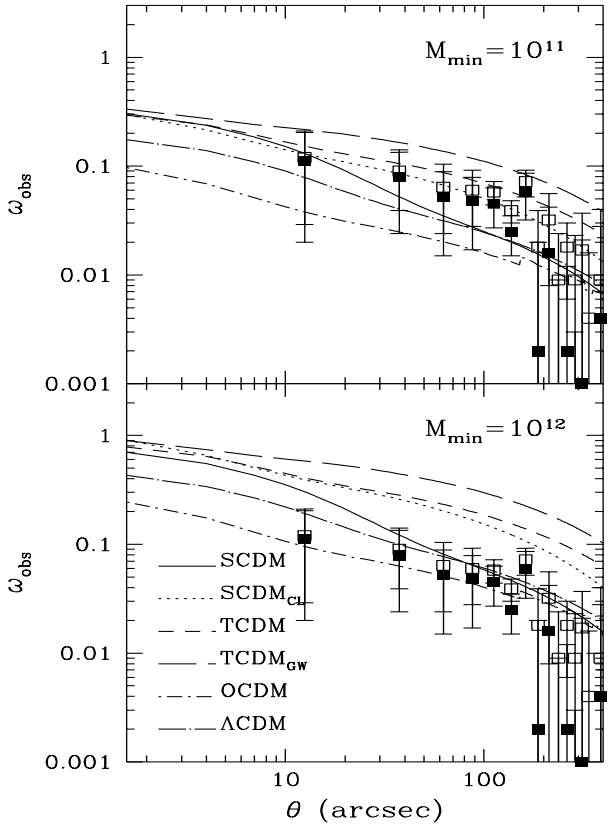,width=9.0cm,height=12.0cm}}  
\caption{ 
Theoretical prediction in different cosmological models for the angular
correlation function of the Lyman--break galaxies. The adopted redshift
distribution and the correlation data are  taken from Giavalisco et al. (1998).
Open and filled squares (with 1$\sigma$ errorbars) refer to two different
estimators of the angular correlation, PB and LS respectively (see Giavalisco
et al. 1998 for a discussion). Two different minimum masses are used to compute
the effective bias: $10^{11} h^{-1} M_\odot$ (top panel) and $10^{12} h^{-1}
M_\odot$ (bottom panel). 
}

\end{figure}

Predictions of the {\em projected} real--space correlation function $w_{\rm
obs} $ can be directly obtained by $\xi_{\rm obs}(r)$ as 
\be
w_{\rm obs}(r_p) = 
2 \int_{r_p}^\infty d r ~r ~(r^2 - r_p^2)^{-1/2}
~\xi_{\rm obs}(r) \;,
\ee
where $r_p$ is the component of the pair separation perpendicular to the line
of sight. Predictions for the projected correlation function are shown in
Figure 2. 

\begin{figure}                  
\centerline{\psfig{file=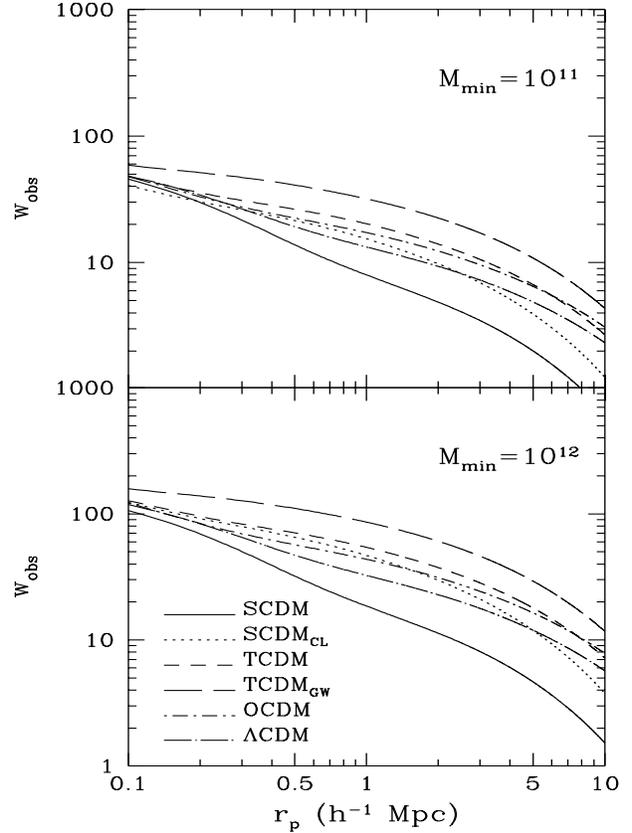,width=9.0cm,height=12.0cm}}  
\caption{ 
Theoretical prediction in different cosmological models for the projected
correlation function of the Lyman-break galaxies as a function of the
(comoving) separation $r_p$ (in units of $h^{-1}$ Mpc). The redshift
distribution is given by Giavalisco et al. (1998). Two different minimum masses
are used to compute the effective bias: $10^{11} h^{-1} M_\odot$ (top panel)
and $10^{12} h^{-1} M_\odot$ (bottom panel). 
}
\end{figure}

Notice that the order of the amplitudes of these curves for different models is
different from the angular correlation function. This is because of the
different weighting by redshift and the dependence on background cosmology of
the formulae. 

One has to be a little cautious about the interpretation of these results
because of the relatively small physical scales being probed by the sky
correlations observed. The formula (4) was derived using quasi-linear arguments
which are not strictly valid on small length scales. In particular, one would
expect that for spatial separations of order half the initial Lagrangian radius
of the haloes, their  correlation function should become negative due to
exclusion effects (Lacey \& Cole 1994; Porciani et al. 1998). This problem is
not restricted to this analysis, but is endemic in studies of this kind (e.g.
Baugh et al. 1998). In practical terms it is particularly relevant in the case
of an Einstein-de Sitter model and for a large halo mass (and therefore large
Lagrangian radius): if $M=10^{12} h^{-1} M_\odot$ then equation (4) is formally
suspect on length scales subtending an angle of about 20 arcseconds at $z=3$ if
$\Omega_0=1$. Recent studies (e.g. Mo \& White 1996) seem to show that the
extrapolation of equation (4) onto small scales is in fairly good agreement
with numerical experiments that incorporate non-linear gravitational effects on
small scales that are absent from the simple theoretical calculation discussed
above. We are therefore fairly confident about these predictions, given the
limitations of the galaxy formation models, but this issue should be explored
more fully using high resolution N-body simulations. 

\section{Discussion} 

The most important results of this work are shown in Figure 1. This shows
clearly that the angular clustering of LBGs is a potential powerful
discriminator between cosmological structure formation models. Different models
predict clearly different angular correlation functions for the LBG
distribution function. The most significant cause of uncertainty is the minimum
halo mass for formation of an LBG. If the critical halo mass is relatively
small, e.g. $10^{11} h^{-1} M_\odot$, then present data do not clearly
discriminate between the suite of models we discuss. On the other hand, if the
minimum mass is a factor of ten larger then only the $\Lambda$CDM and OCDM
models are allowed. In such a case the data favour a Universe with low
matter-density, although they do not appear to be strongly sensitive to the
presence or absence of a vacuum energy density (i.e. a cosmological constant
term). 

It is clearly important to decide what value of the threshold mass is more
appropriate for this population of objects. Theoretical arguments generally
favour the higher of the two values we have suggested here (e.g. Steidel et al.
1998; Paper II). Observationally, one might hope to determine the masses of the
LBGs through their velocity dispersions. But the likeliest explanation of these
objects at present seems to be that they form within rather massive haloes, but
do not yet correspond to fully-assembled galaxies. Velocity dispersions might
therefore merely reflect the local velocities of star-forming regions, which
would lead to an underestimate of the halo mass. Such data would therefore need
to be interpreted within the framework of a complete model of galaxy formation
and evolution (Baugh et al. 1998; Governato et al. 1998). Such models would
also establish a clearer connection the LBG population and that of nearby
galaxies at $z\simeq 0$. 

Despite the uncertainties surrounding the nature and identity of the LBGs, it
is reassuring that their clustering properties nevertheless seem to fit within
the standard gravitational instability scenario. Moreover, they also seem to be
in line with the emerging consensus that we live in a Universe with a matter
density which is less than the critical density. 

\section*{Acknowledgments.} 

PC is a PPARC Advanced Research Fellow. He is grateful to the Dipartimento di
Astronomia at the Universit\`{a} di Padova for hospitality during a visit when
this work was begun. The Italian MURST is acknowledged for partial financial
support. We all thank Mauro Giavalisco for sending us the observational data
pertaining to LBGs.

\end{document}